\documentclass[aps, prc, twocolumn, superscriptaddress, nofootinbib]{revtex4-1}
\usepackage{graphicx}
\usepackage[english]{babel}
\usepackage{bm}
\usepackage[colorlinks=true, allcolors=blue, pdfusetitle]{hyperref}

\newcommand{\subfigimg}[3][,]{%
  \setbox1=\hbox{\includegraphics[#1]{#3}}
  \leavevmode\rlap{\usebox1}
  \rlap{\hspace*{35pt}\raisebox{\dimexpr\ht1-3\baselineskip}{#2}}
  \phantom{\usebox1}
}

\begin{document}

\title{Collision Geometry and Flow in Uranium+Uranium Collisions}

\author{Andy Goldschmidt}
\affiliation{Department of Physics, The Ohio State University, Columbus, OH 43210, USA}
\author{Zhi Qiu}
\affiliation{Department of Physics, The Ohio State University, Columbus, OH 43210, USA}
\author{Chun Shen}
\affiliation{Department of Physics, McGill University, Montreal, Quebec, H3A 2T8, Canada}
\affiliation{Department of Physics, The Ohio State University, Columbus, OH 43210, USA}
\author{Ulrich Heinz}
\affiliation{Department of Physics, The Ohio State University, Columbus, OH 43210, USA}

\begin{abstract}
Using event-by-event viscous fluid dynamics to evolve fluctuating initial density profiles from the Monte-Carlo Glauber model for U+U collisions, we report a ``knee''-like structure in the elliptic flow as a function of collision centrality, located around the 0.5\% most central collisions as measured by the final charged multiplicity. This knee is due to the preferential selection of tip-on-tip collision geometries by a high-multiplicity trigger. Such a knee structure is not seen in the STAR data. This rules out the two-component MC-Glauber model for initial energy and entropy production. Hence an enrichment of tip-tip configurations by triggering solely on high-multiplicity in the U+U collisions does not work. On the other hand, by using the Zero Degree Calorimeters (ZDCs) coupled with event-shape engineering such a selection is possible. We identify the selection purity of body-body and tip-tip events in full-overlap U+U collisions. By additionally constraining the asymmetry of the ZDC signals we can further increase the probability of selecting tip-tip events in U+U collisions. 
\end{abstract}

\maketitle

\section{Introduction}
\label{sec:1}

High energy collisions between heavy ions are used to probe emergent phenomena in Quantum Chromodynamics (QCD), the theory of the strong interaction.  One feature of QCD is the transition from hadronic matter to a color-deconfined quark-gluon plasma (QGP) \cite{Heinz:2002gs, Gyulassy:2004vg, Shuryak:2004cy} as the temperature is increased. This transition can occur in heavy-ion collisions with sufficient energy for the system to melt into a hot dense fireball of asymptotically free quarks and gluons.

Relativistic hydrodynamic models have been successful in describing the dynamical evolution of QGP \cite{Kolb:2003dz}. Motivated as a testing ground for these models, a U+U collisions program was recommended in order to study the unique collision geometry resulting from the prolate deformation of the uranium nucleus \cite{Shuryak:1999by, Kolb:2000sd, Heinz:2004ir, Kuhlman:2005ts, Nepali:2006ep, Nepali:2007an}. Such a program was carried out in 2012 at the Relativistic Heavy Ion Collider (RHIC) at Brookhaven National Lab \cite{Wang:2014qxa}.

To understand the attraction of uranium, consider that the initial temperature distribution of each QGP droplet is controlled by two main factors: deterministic collision geometry (i.e. the shape of the overlap region between two nuclei), and quantum mechanical fluctuations in the nucleon positions. For spherical nuclei, the collision geometry is entirely a function of the impact parameter.  However, in prolate deformed uranium, the geometry of the initial temperature distribution also depends on the relative spatial orientation of the two nuclei which can be described by the Euler angles between their long major axis.

We focus in this paper on two limiting cases for fully overlapping uranium collisions. At one extreme are the ``tip-tip'' events, defined to occur when the major axes of the nuclei lie parallel with the beam direction. Residing at the opposite limit are the ``body-body'' events, where the major axis of each nucleus is perpendicular to the beam direction. We are interested in answering the question of how and with what precision we can distinguish experimentally between these configurations. Their conceptual importance is explained in \cite{Heinz:2004ir}.

A preliminary account of parts of this work can be found in \cite{Goldschmidt:2015qya}; however, all results shown in Sec.~\ref{sec3b} involving hydrodynamically evolved events with ZDC selection constraints are new, and some of the results presented in Sec.~\ref{sec3a} are based on larger hydrodynamic event samples. 

\section{The model}
\label{sec:2}

To model the initial energy density distribution of U+U collisions we employ the two-component (wounded nucleon/binary collisions) Monte-Carlo Glauber model. We use the deformed Woods-Saxon \cite{Woods:1954zz} distribution
\begin{equation}
 \rho(r, \theta, \varphi) = \frac{\rho_0}{1 + e^{(r - r(\theta, \varphi))/d}}
 \label{eq1}
\end{equation} 
to sample the positions of nucleons inside a uranium nucleus.  In Eq.~(\ref{eq1}) we take for the surface diffusiveness parameter $d=0.44 \mathrm{\ fm}$ and for the saturation density parameter $\rho _0= 0.1660 \mathrm{\ fm}^{-3}$ \cite{Chamon:2002mx, Hirano:2010jg}. The spatial configuration of a uranium nucleus is deformed; we model its charge radius as \cite{Moller:1993ed}
\begin{equation}
r(\theta ,\varphi ) = r_{ 0 }( 1+\sum_{ l=1 }^{ \infty }{ \sum_{ m=-l }^{ l } \beta_{ lm }Y_{ l }^{ m }(\theta, \phi) }),
\label{eq2}
\end{equation}
where $r_0=6.86 \mathrm{\ fm}$ is the average nuclear charge radius \cite{Hirano:2010jg}. We assume the uranium nucleus is azimuthally symmetric and choose the non-vanishing deformation parameters $\beta_{20}=0.28$ and $\beta_{40}=0.093$ for the quadrupole and hexadecupole deformations along its main axis \cite{Filip:2009zz}. The choices of these parameters agree well with  a recent reanalysis in \cite{Shou:2014eya}, except for $\beta_{20}$ for which \cite{Shou:2014eya} gives the value $0.265$.

We use the Woods-Saxon density (\ref{eq1}) to Monte-Carlo sample the nucleon centers and represent each nucleon in the transverse plane by a gaussian areal density distribution about its center:
\begin{equation}
 \rho_n(\vec{\mathbf{r}}_{\perp}) = \frac{1}{2 \pi B}e^{-r_\perp^2/(2B)} .
 \label{eqn3}
\end{equation}
The width parameter $B = \sigma_\mathrm{NN}^{in}(\sqrt{s_\mathrm{NN}})/14.30$ depends on collision energy as described in \cite{Heinz:2011mh}.  

The sum of these gaussian nucleon density distributions represents the nuclear density distribution for the sampled nucleus at the time of impact and is used to compute the initial energy density distribution generated in the collision.  For this calculation, we use the two-component Monte-Carlo Glauber model which adds contributions from binary collisions $N_b$ and wounded nucleon participants $N_p$ with an adjustable relative weight \cite{Bialas:1976ed}. The collision criterium that identifies binary collisions and wounded nucleons is evaluated probabilistically with the Gaussian nucleon profile (\ref{eqn3})  \cite{Heinz:2011mh}.

The binary collision term counts the entropy deposited by pairs of colliding nucleons and is modeled by a gaussian distribution with the same size as a nucleon (see Eq.~(\ref{eqn3}) \cite{Shen:2014vra}; the total binary collision density per unit transverse area is
\begin{equation}
 n_{BC} (\vec{\mathbf{r}}_{\perp}) = \sum _{ i,j }{ \gamma_{i,j} \frac{1}{2\pi B}e^{-\left| \vec{ \mathbf{ r}}_{\perp} - \vec{ \mathbf{ R}}_{ i,j }  \right| ^2  / (2B)}}
  \label{eqn4}
\end{equation}
where the sum is over all pairs of colliding nucleons. The normalization $\gamma_{i}$ is a $\Gamma$-distributed random variable with unit mean that accounts for multiplicity fluctuations in individual nucleon-nucleon collisions.

Each struck nucleon is said to be wounded by (or participating in) the collision and contributes a portion of the initial entropy density distributed symmetrically about its center; the resulting total wounded nucleon density per unit area is given by
\begin{equation}
 n_{WN} (\vec{\mathbf{r}}_{\perp}) = \sum_{ i }{ \gamma_{i} \frac{1}{2\pi B}e^{-\left| \vec{ \mathbf{ r}}_{\perp} - \vec{ \mathbf{ r}}_{ i, \perp }  \right| ^2  / (2B)}}
 \label{eqn5}
\end{equation}
where the sum is over all wounded nucleons in both nuclei and $\gamma_{i}$ is again a fluctuating factor with unit mean.

As mentioned, we model multiplicity fluctuations in single nucleon-nucleon collisions using $\gamma_{i,j}$ and $\gamma_{i}$, which are taken to be $\Gamma$-distributed random variables with unit mean and and with variances controlled by parameters $\theta_{BC}$ and $\theta_{WN}$, respectively. The generic $\Gamma$ distribution with unit mean and scale parameter $\theta$ is given by:
\begin{equation} \label{4.1}
\Gamma \left( \gamma ; \theta  \right) = \frac { \gamma^{ 1/\theta - 1 }{ e }^{ - \gamma / \theta } }{ \Gamma \left( 1/\theta  \right) { \theta  }^{ 1/\theta  }  } { , }\quad \gamma \in \left[ 0, \infty  \right)
 \label{eqn6}
\end{equation}
The multiplicity fluctuations from wounded nucleons and binary collisions are related by requiring \cite{Shen:2014vra}:
\begin{equation}
 \theta_{pp} = \frac{ 1-\alpha }{2} \theta_{WN}  = \alpha \theta_{BC} .
  \label{eqn7}
\end{equation}
The parameter $\theta_{pp} = 0.9175$ was fit to multiplicity distributions measured in p+p collisions, assuming multipli\-city fluctuations coming purely from the initial state \cite{v1}.

The distribution in the transverse plane of the deposited entropy per unit volume is determined by mixing the binary collision and wounded nucleon sources using
\begin{equation}
s_{ 0 }(\vec { \mathbf{r}}_{ \perp  })=\frac { \kappa _{ s } }{ \tau_{ 0 } } \left( \frac { 1-\alpha  }{ 2 } n_{ { WN } }(\vec { \mathbf{r} }_{ \perp  })+\alpha n_{ { BC } }(\vec { \mathbf{r} }_{ \perp  }) \right) .
\label{eq8}
\end{equation}
where $\tau_0$ is the starting time for the (hydro)dynamical evolution of the collision fireball. We choose $\kappa_s = 17.16$ and the mixing ratio $\alpha=0.12$ to reproduce the measured charged multiplicities and their dependence on collision centrality in Au+Au collisions at 200 $A$\,GeV. The shape of the resulting energy density distribution in the transverse plane is calculated from the entropy density using the equation of state (EoS) \verb|s95p-v0-PCE| from Lattice QCD \cite{Huovinen:2009yb}.  The initial energy profile is evolved using the viscous relativistic fluid dynamic code package \verb|iEBE-VISHNU| \cite{Shen:2014vra} with specific shear viscosity $\eta/s = 0.08$. Simulations begin at time $\tau_0 = 0.6 \ \mathrm{fm/c}$ and decouple at a temperature $T_{ \mathrm{dec} }= 120$\,MeV. The single particle momentum distribution is then computed using the Cooper-Fyre Formula. A full calculation of charged hadron observables that includes all hadronic resonance decay processes on an event-by-event basis is numerically costly; for this reason we computed only the directly emitted positively charged ``thermal pions'', $\pi^+$ and take this quantity as a measure for total charged multiplicity.  At a fixed freeze-out temperature of 120 MeV, the two quantities are related by a constant factor: $dN_{ch}/d\eta \simeq 4.6 \ dN_{\pi^+}/dy$.

The initial energy density profiles fluctuate from event to event. Each profile can be characterized by the $r^n$-weighted eccentricity coefficients $\varepsilon_n$ and their associated ``participant plane angles'' $\Phi_n$:
\begin{equation}
\mathcal{E}_n := \varepsilon _{ n }e^{ in\Phi _{ n } }=-\frac { \int  d\vec{\mathbf{r}}_{\bot} r^{ n }e^{ in\varphi  }e(\vec{\mathbf{r}}_{\perp}) }{ \int  d\vec{\mathbf{r}}_{\bot} r^{ n }e(\vec{\mathbf{r}}_{\perp}) }\qquad (n \ge 2) . 
\label{eq9}
\end{equation}
where $(r, \varphi)$ are the standard polar coordinates in the transverse plane and $e(\vec{\mathbf{r}}_{\perp})$ is the initial energy density \cite{Alver:2010gr}.
Through the hydrodynamic evolution, these spatial eccentricities $\{\epsilon_n,\Phi_n\}$ translate themselves into the anisotropic flow coefficients $\{v_n, \Psi_n\}$ \cite{Gardim:2011xv, Voloshin:1994mz, Qiu:2011iv}:
\begin{equation}
\mathcal{V}_n := v_n e^{in\Psi_n}=\frac{ \int  p_T dp_T d \varphi_p e^{i n \varphi_p} dN/ (p_T dp_T d \varphi_p) }{ \int  p_T dp_T d \varphi_p dN/ (p_T dp_T d \varphi_p)} .
\label{eq10}
\end{equation}

Apart from the Monte-Carlo Glauber model, there exist various other initialization models. These include the IP-Glasma model \cite{Schenke:2012wb}, the MC-KLN model \cite{Kharzeev:2000ph,Kharzeev:2002ei, Drescher:2006ca, Drescher:2007ax}, and the TRENTO model \cite{Moreland:2014oya}. As we will see, U+U collisions can provide experimental measurements to distinguish between these various initializations.

\begin{figure*}
  \begin{tabular}{cc}
  \begin{tabular}{cc}
  \includegraphics[width = 0.5\textwidth]{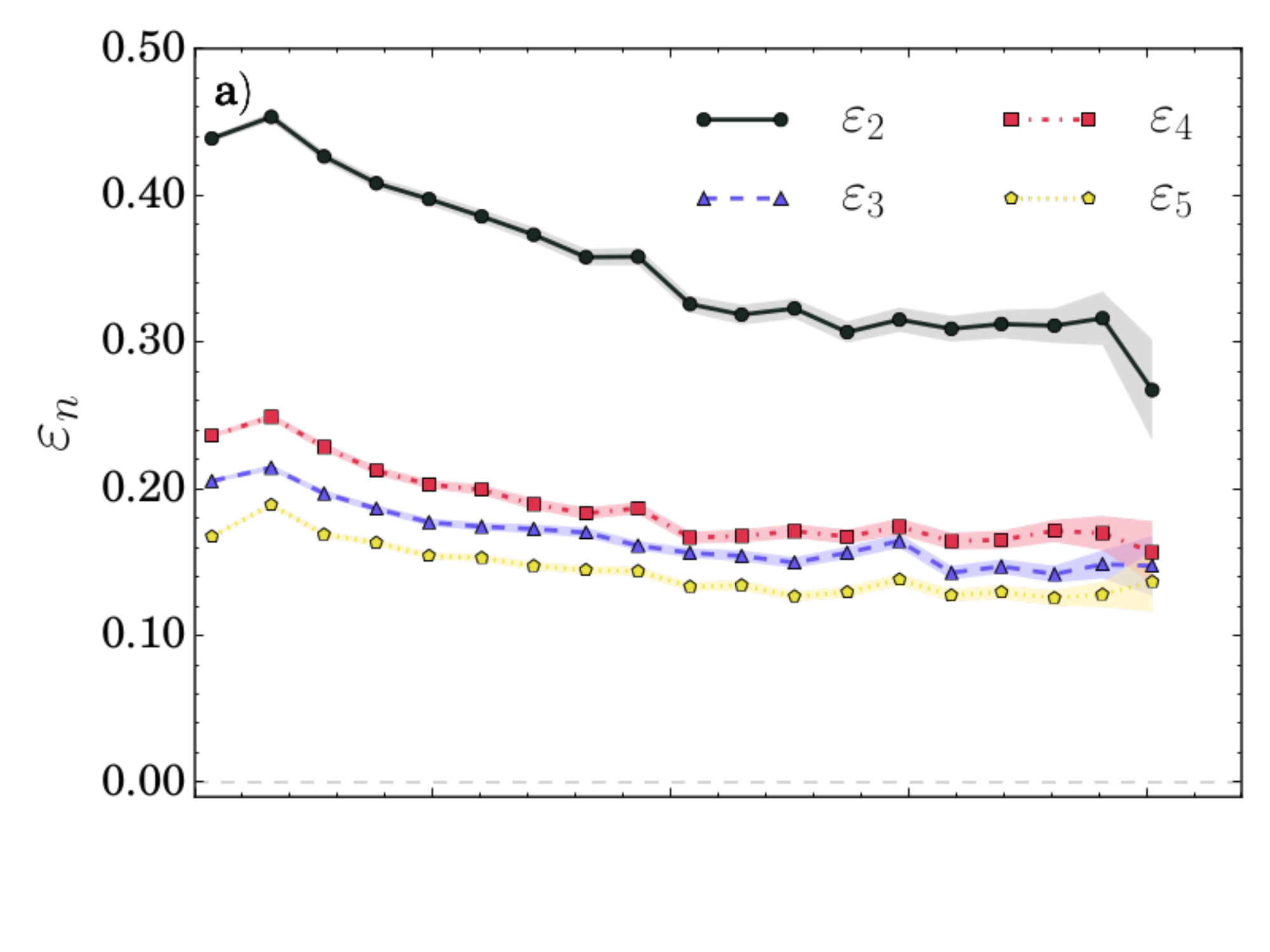}\hspace{-.75em}%
  \includegraphics[width = 0.5\textwidth]{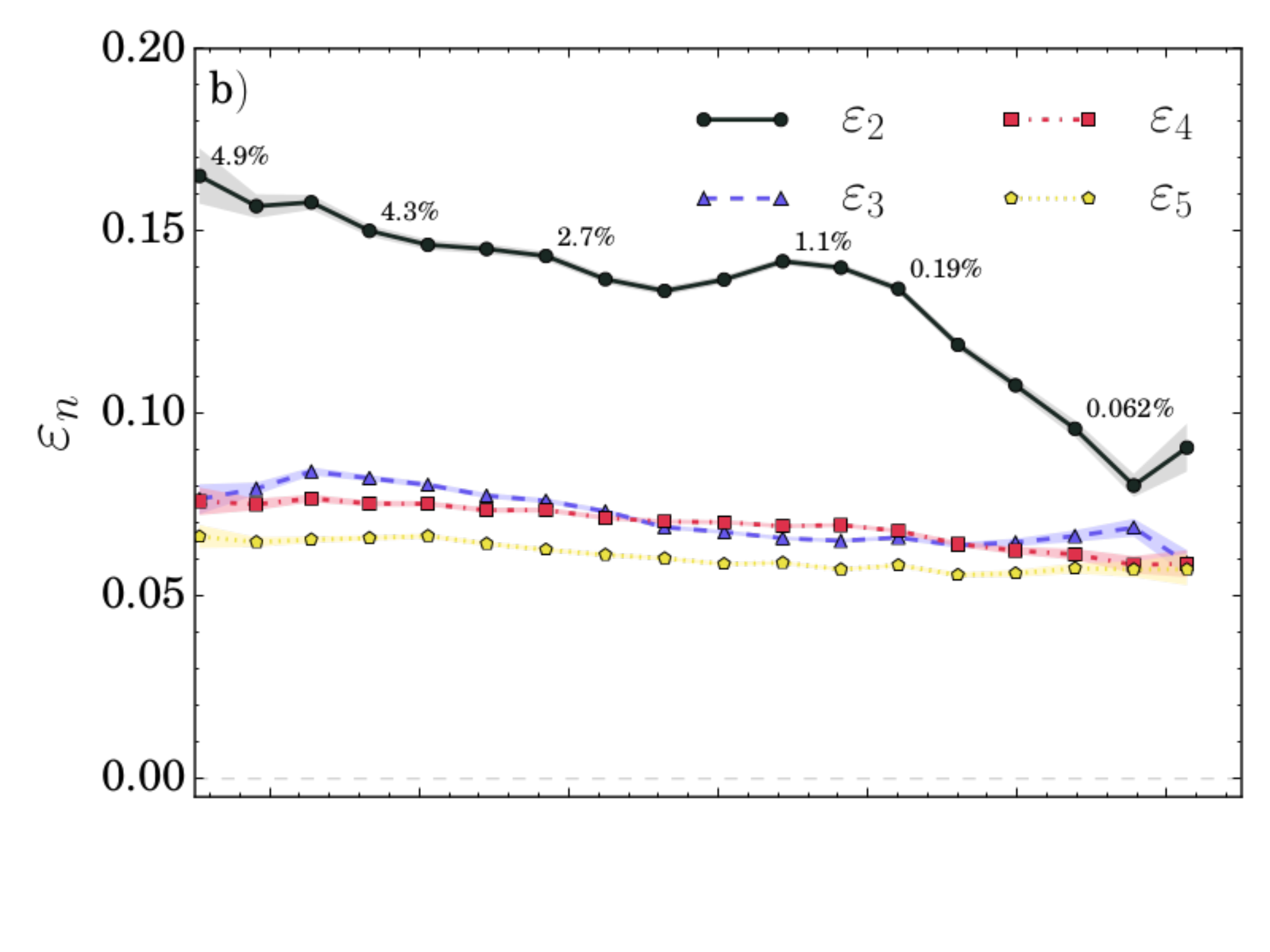} \\ [-10.25ex]
  \includegraphics[width = 0.5\textwidth]{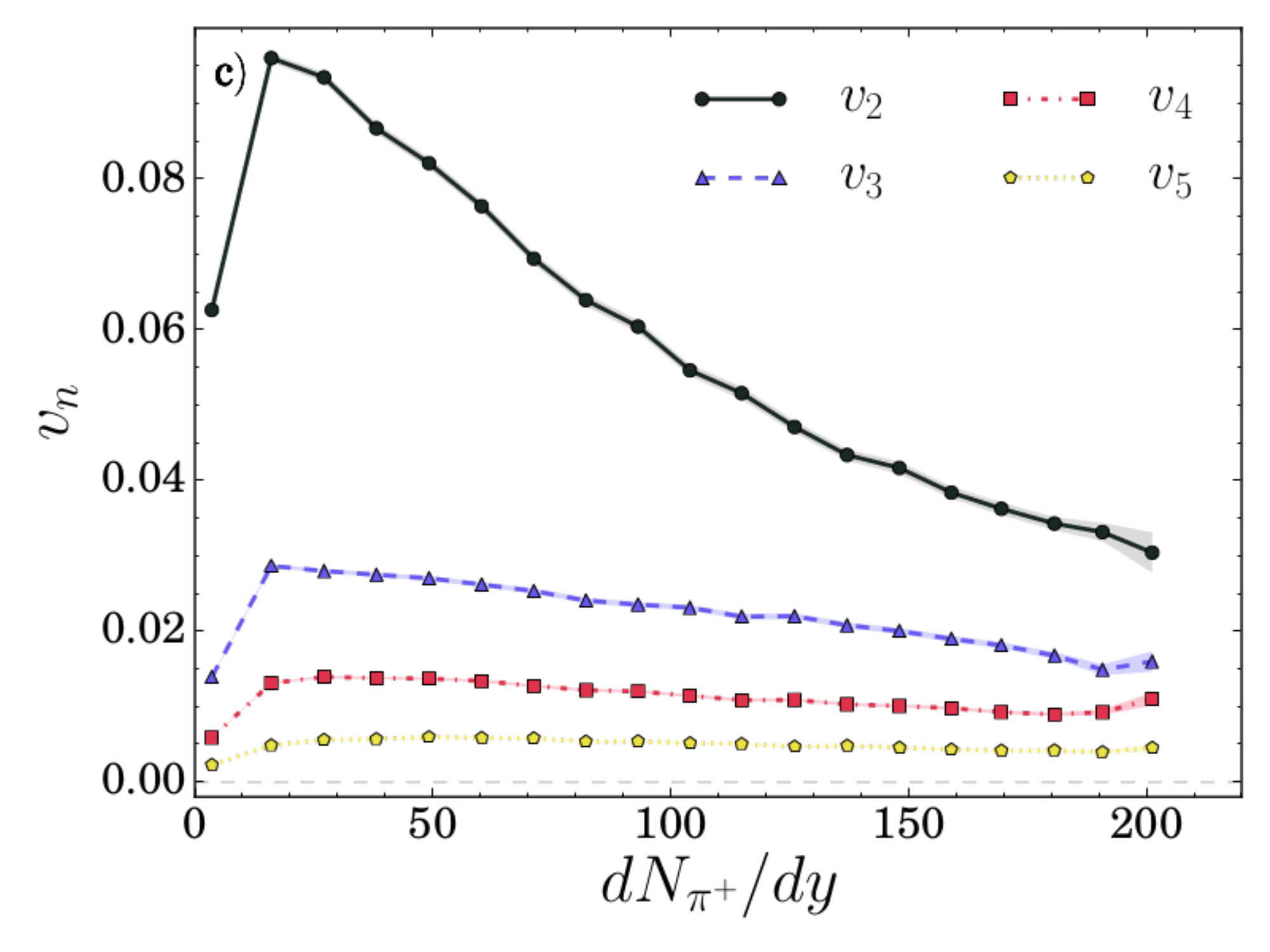}\hspace{-.75em}%
  \includegraphics[width = 0.5\textwidth]{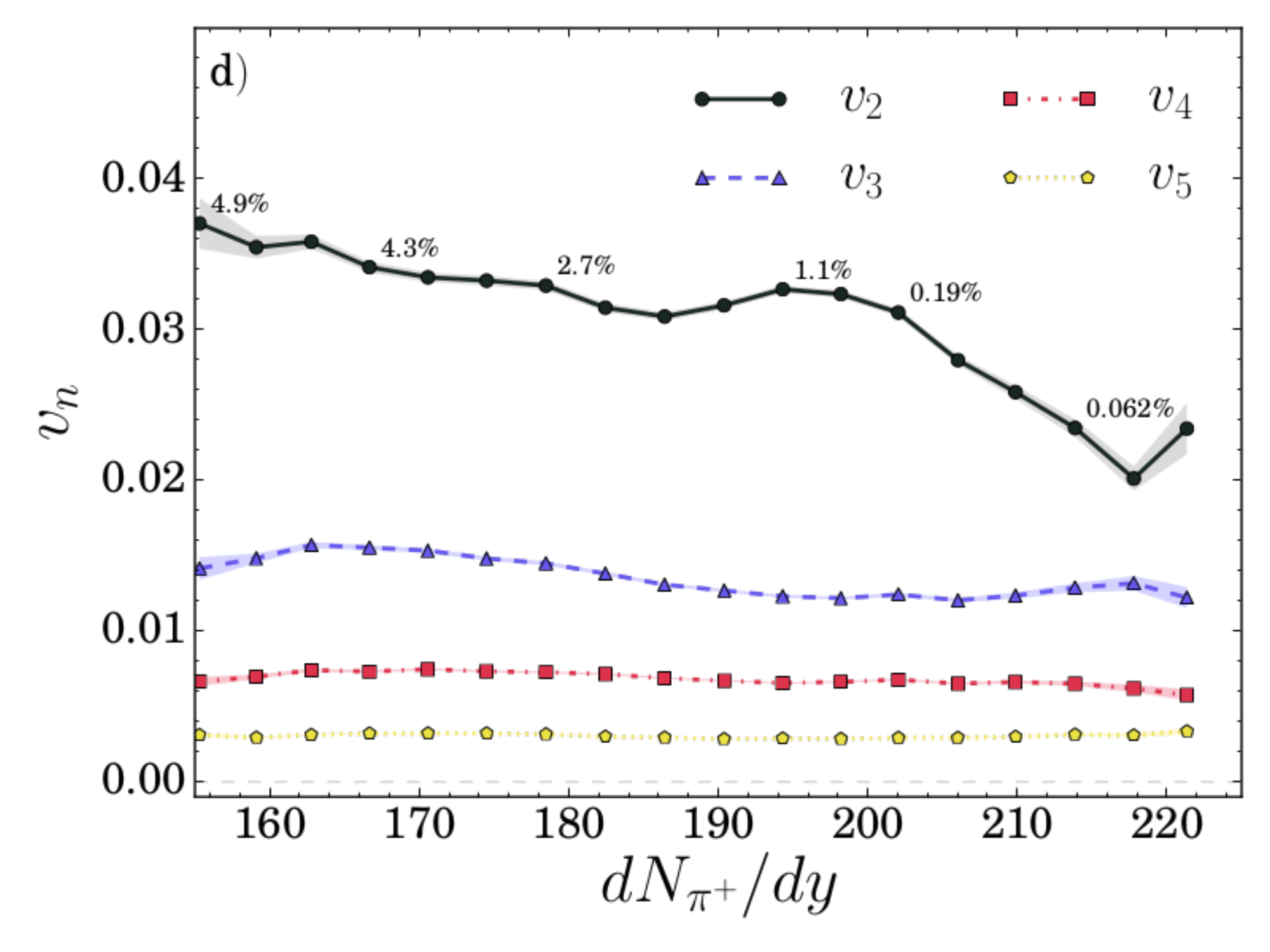} \\[-3ex]
  \end{tabular}
  \end{tabular}
  \caption{Panels (a,b) show the event-averaged eccentricities $\varepsilon_n$, before hydrodynamic evolution, panels (c,d) the event-averaged flows $v_n$ after hydrodynamic evolution. The left panels (a,c) represent 35,000 minimum bias events that include multiplicity fluctuations whereas the right panels (b,d) were obtained from a different set of 35,000 multiplicity-selected events, generated earlier without including multiplicity fluctuations, covering only the 0-5\% centrality range.}
  \label{fig:2}
\end{figure*}

\section{Constraining collision geometry with multiplicity, flow, and ZDC cuts}
\label{sec:3}
\subsection{Eccentricity and flow coefficients as a function of multiplicity}
\label{sec3a}

In Fig.~\ref{fig:2}, we present the centrality dependence of the initial eccentricities and the final anisotropic flow coefficients of thermal pions for harmonic order $n = 2 - 5$ in U+U collisions at 193 $A$\,GeV. In Figs.~\ref{fig:2}a,c minimum bias results are shown as functions of the thermal pion yields, $dN_{\pi^+}/dy$. We notice that the variance of $\varepsilon_{2,4}$ and $v_{2,4}$ in ``most central'' (i.e. highest multiplicity) collisions are larger than in the rest of the centrality range.  This is because in the most central collisions the two uranium nuclei are colliding with impact parameter $b \approx 0$ but, as a result of the large spatial deformation, not always with full overlap. A mixture of tip-tip and body-body collisions in these highest multiplicity events increases the variance of the initial $\varepsilon_{2,4}$ which in turn drives a larger variance in $v_{2,4}$.

In Figs.~\ref{fig:2}b,d we focus on the 0-5\% most central U+U collisions and increase the statistics to 35,000 events for just this bin. We find a ``knee'' structure in the high multiplicity regime ($<0.5\%$ centrality) for both $\varepsilon_2$ and $v_2$. This can be understood as follows: First, while the ellipticity in the transverse plane for a tip-tip collision is small (as the overlap area is approximately circular), body-body collisions produce ellipsoidally deformed ovelap regions with larger ellipticities $\varepsilon_2$.  Second, although fully overlapping tip-tip and body-body collisions share the same number of participants, more binary collisions between nucleons can happen in the optically thicker tip-tip event, implying (in our two-component Glauber model) a larger initial $dS/dy$ deposited for the tip-tip configuration. In the presence of fluctuations which lead to a range of $\varepsilon_2$ values for a given $dS/dy$ and vice-versa, the larger average multiplicity in tip-tip collisions implies an increasing bias toward small $\varepsilon_2$ when selecting events with larger and larger values of $dS/dy$. This preferential selection of tip-tip orientations at high multiplicities accounts for the appearance of a knee structure in the initial ellipticity \cite{Voloshin:2010ut} (Fig.~\ref{fig:2}).  We see in Fig.~\ref{fig:2}c that the knee is preserved after an event-by-event hydrodynamic simulation when plotting the elliptic flow of the final particle distribution as a function of multiplicity. 

We emphasize that that experimental results from STAR do not show this knee structure \cite{Wang:2014qxa}. Considering the preservation of the structure after hydrodynamic evolution as seen in Fig.~\ref{fig:2}, we conclude that, in contrast to Au+Au collisions where it has been extensively tested, the two component MC-Glauber model fails to correctly identify entropy production in ultra-central U+U collisions where the knee is predicted by the model but not found experimentally. Hence the non-linear dependence of multiplicity on the number of wounded nucleons observed in spherical Au+Au and Pb+Pb collisions as a function of collision centrality cannot be attributed to a binary collision component as implemented in the two-component MC-Glauber model. 

The IP-Glasma model \cite{Schenke:2014tga} and the wounded-constituent-quark Glauber model \cite{Adler:2013aqf} provide alternative explanations of the measured centrality dependence of the charged multiplicity without invoking a binary collision component. Gluon saturation physics as implemented in the IP-Glasma model is able to simultaneously accomodate a strong nonlinearity of $dN/dy$ as a function of $N_\mathrm{part}$ in Au+Au and Pb+Pb collisions and a weak dependence of $dN/dy$ on collision orientation in central U+U collisions with a fixed number of participants, while the MC-Glauber model cannot \cite{Schenke:2014tga}. Alternatively, it has been suggested in \cite{Adler:2013aqf} that a model that produces entropy according to the number of wounded valence gluons (rather than wounded nucleons) can also reproduce the observed {\it nonlinearity} of $dN/dy$ as a function of participant nucleons in Au+Au and Pb+Pb at RHIC and LHC, without a binary collision component. It would be interesting to study the prediction of such a model for central U+U collisions of varying orientations.

\begin{figure}
\centering
  \includegraphics[width = 0.5\textwidth]{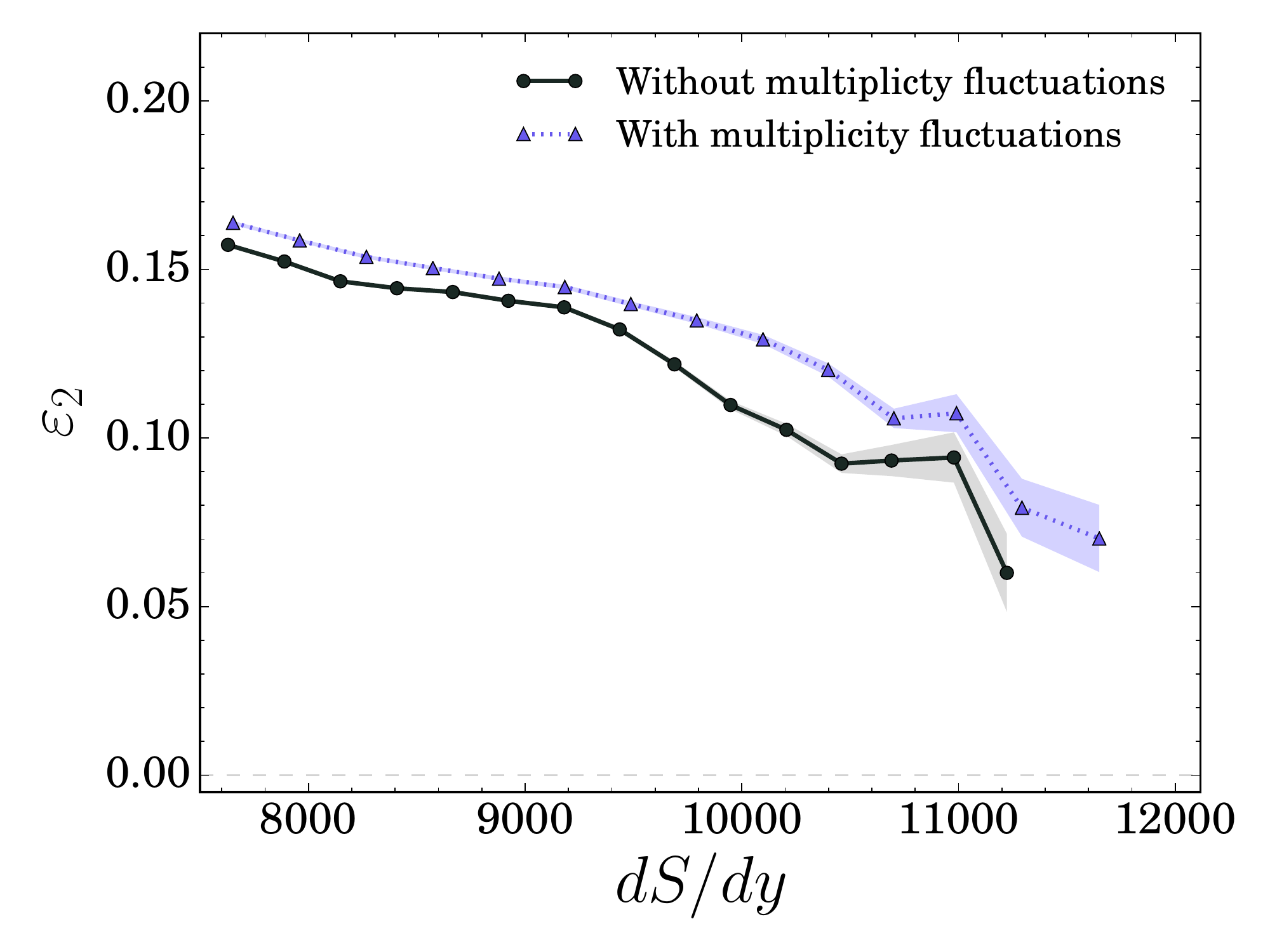}
  \caption{The ellipticity $\varepsilon_2$ as a function of $dS/dy$ from the MC-Glauber model, for collisions roughly in the 0-5\% centrality range, with (blue dashed line) and without (black solid line) multiplicity fluctuations from single p+p collisions.}
  \label{fig:3}
\end{figure}

Some corrections to the entropy production in these ultra central events arise from the inclusion of p+p multiplicity fluctuations. In Fig.~\ref{fig:3} we show that accounting for multiplicity fluctuations increases the average ellipticity of the collision fireball and weakens but does not erase the knee structure in $\varepsilon_2$ vs. $dS/dy$. Hence, this effect alone does not appear sufficient to reach agreement of the MC-Glauber model with data for ultra-central U+U collisions.  We acknowledge that more drastic fluctuation models \cite{Rybczynski:2012av} have been suggested in order to more successfully adjust the theoretical predictions of MC-Glauber to experimental results.

\subsection{Selecting high overlap events with combined ZDC and multiplicity cuts}
\label{sec3b}

In an experimental analysis of relativistic heavy ion collisions, the charged hadron multiplicity, $dN_\mathrm{ch}/dy$ and the elliptic flow coefficient $v_2$ can be used to classify events. Hydrodynamic studies have shown that the initial $\varepsilon_2$ maps linearly to the $v_2$ of hadrons \cite{Qiu:2011iv} and that the initial $dS/dy$ is monotonically related to the final total particle multiplicity, $dN/dy$ \cite{Shen:2014vra}.  Hence one should be able to use $dS/dy$ and $\varepsilon_2$ from the initial conditions as proxies for the (computationally intensive) charged hadron $dN_\mathrm{ch}/dy$ and $v_2$ when testing our ability to select the fully overlapping tip-tip and body-body U+U collisions. In this subsection we test this idea.

In our analysis we make theoretical approximations for the use of experimental forward and backward zero degree calorimeters (ZDCs). Placed at zero degrees far from the colliding pair, ZDCs catch information about the spectator neutrons that pass through a collision without getting wounded. We classify our collisions by using the number of spectators $N_s = 476 - N_\mathrm{part}$ (where $N_\mathrm{part}$ is computed from the MC Glauber model) to mimic the experimental ZDC signal \cite{Kuhlman:2005ts}. For this study we look at 65,000 intial condition events in the $1\%$ most participating ZDC range ($N_s < 19$) and compare statements made on the basis of their ellipticities $\varepsilon_2$ and initially produced entropy $dS/dy$ with analogous statements made on the basis of elliptic flow $v_2$ and final charged multiplicity $dN_\mathrm{ch}/dy$ for a subset of 25,000 events that were evolved using viscous relativistic fluid dynamics. The calculations include $\Gamma$-distributed multiplicity fluctuations in nucleon-nucleon collisions.  Selecting the most participating ZDC collisions allows for a restriction of the set of collisions to more fully overlapping events. In such a regime, any initial geometric effects should come almost exclusively from the deformed shape of the uranium nucleus.

We define the tip-tip and body-body event classes using the pair of angles $(\theta_{1,2}, \phi_{1,2})$ from the two incoming nuclei, where $\theta$ is the polar angle between the long major axis of the uranium nucleus and the beam direction and $\phi$ is its azimuthal angle in the transverse plane. An event is classified ``tip-tip'' if $\sqrt{ (\cos^2 \theta_1 + \cos^2 \theta_2)/2 } > \sqrt{3}/2$, and ``body-body'' if both $\sqrt{(\cos^2 \theta_1 + \cos^2 \theta_2)/2} < 1/2$ and $\left|\phi_1 - \phi_2 \right| < \pi/10$.  The polar angle constraints are chosen such that if the first nucleus has $\theta_1 = 0$ the collision is called ``tip-tip'' for $\theta_2 < \pi/4$.  Accordingly, if the first nucleus has $\theta_1 = \pi/2$, the collision is called ``body-body'' for $\theta_2 > \pi/4$.  These polar angle constraints imply that if $\theta_1 = \theta_2$, the common angle is less than $\pi/6$ for ``tip-tip'' and greater than $\pi/3$ for ``body-body''.  The additional azimuthal constraint included for body-body events is intended to force good alignment of the long major axes.

\begin{figure}[h]
  \hspace*{-3mm}\includegraphics[width = 0.55\textwidth]{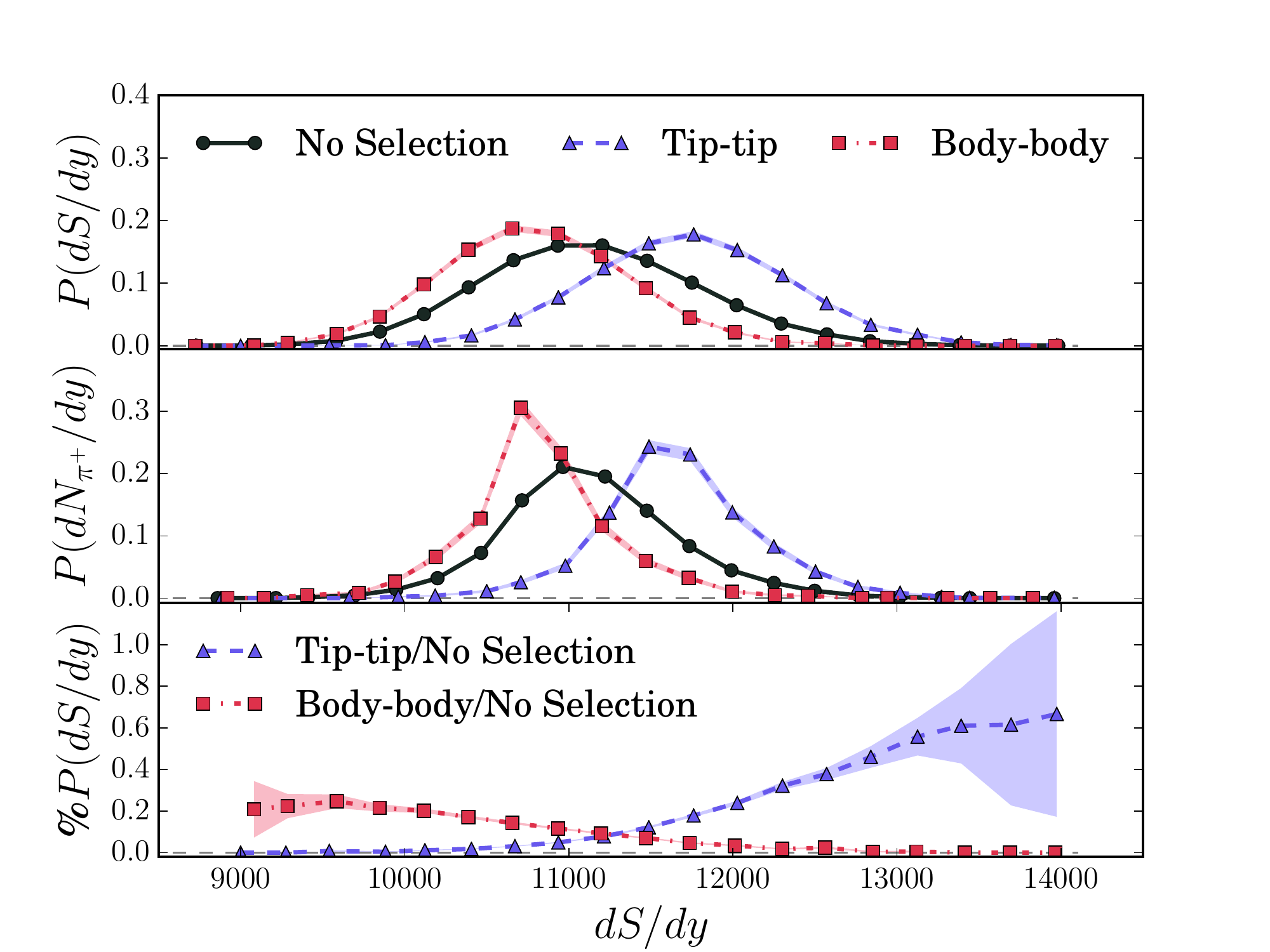}
  \caption{Probability distributions for $dS/dy$ (top) and $dN_{\pi^+}/dy$ (middle, scaled by $\langle dS/dy \rangle / \langle dN_{\pi^+}/dy \rangle$) showing tip-tip and body-body collision contributions, within a selection of 1\% ZDC events. The bottom panel shows the relative probabilities for tip-tip and body-body events among all events of a given $dS/dy$.}
  \label{fig:4}
\end{figure}

In Fig.~\ref{fig:4} we plot the probability distributions for $dS/dy$ and $dN_{\pi^+}/dy$. Using our collision definitions we can directly read from the figure the likelihood of selecting a certain orientation based on a given multiplcity cut.  Noting the evident congruence of the plots for $dS/dy$ (upper) and $dN_{\pi^+}/dy$ (middle), we rely on $dS/dy$ in the bottom plot because of the improved statistics.  We see that by cutting (within our 1\% ZDC sample) on events with large $dS/dy$ we can enrich the fraction of tip-tip events to about 50\%, whereas cutting on low $dS/dy$ enriches the fraction of body-body events, but never to more than about 20\%. The 20\% limit arises from admixtures from imperfectly aligned collisions that are not really "full overlap''.  The enrichment of tip-tip or body-body by varying $dS/dy$ relies on the assumed two-component nature of entropy production which also produced the knee structure discussed before. Indeed, selection efficiency of specific collision geometries by cutting on $dS/dy$ is model dependent.

\begin{figure}
  \subfigimg[width = \linewidth]{a)}{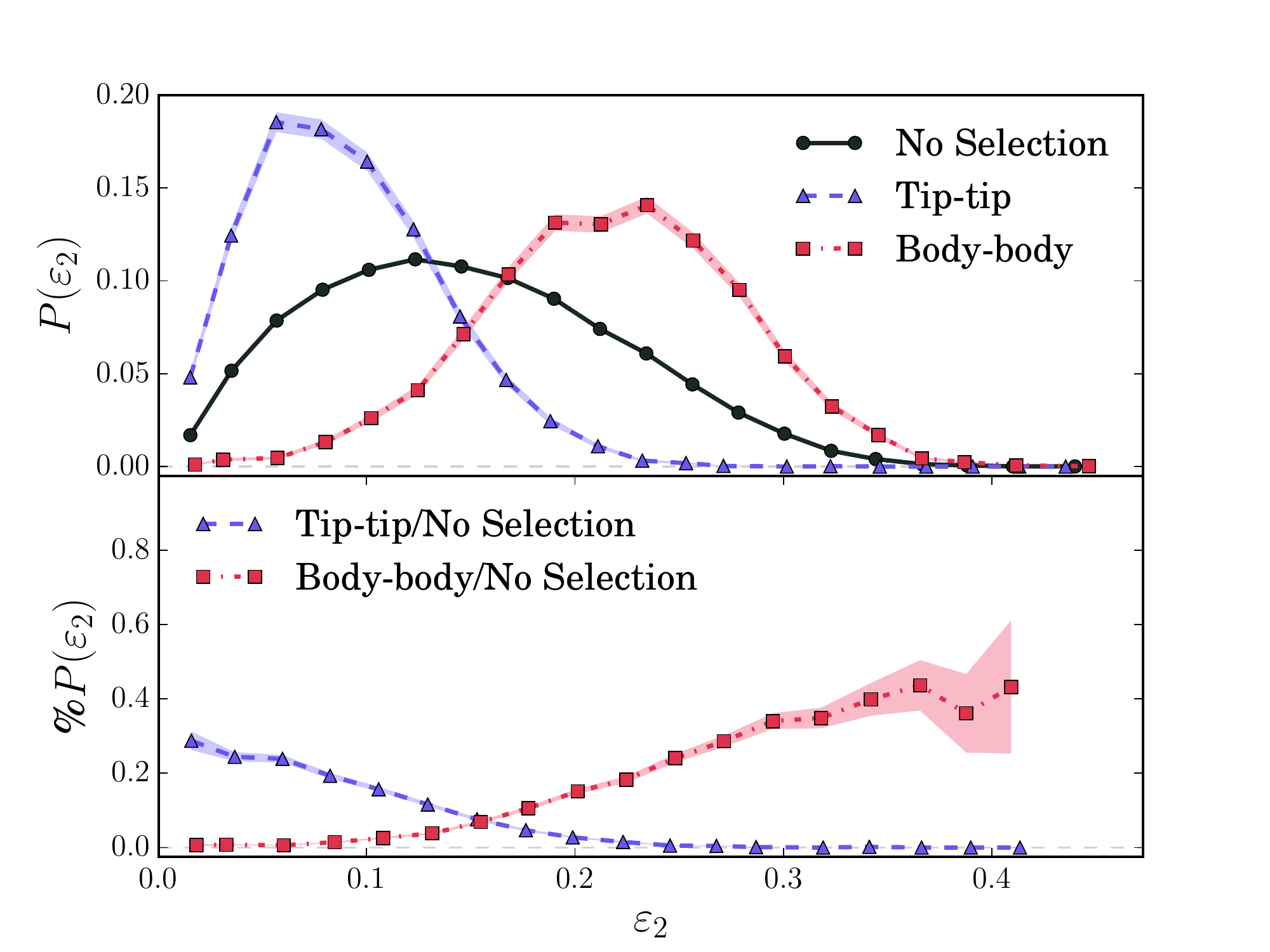}\\[-4ex]
  \subfigimg[width = \linewidth]{b)}{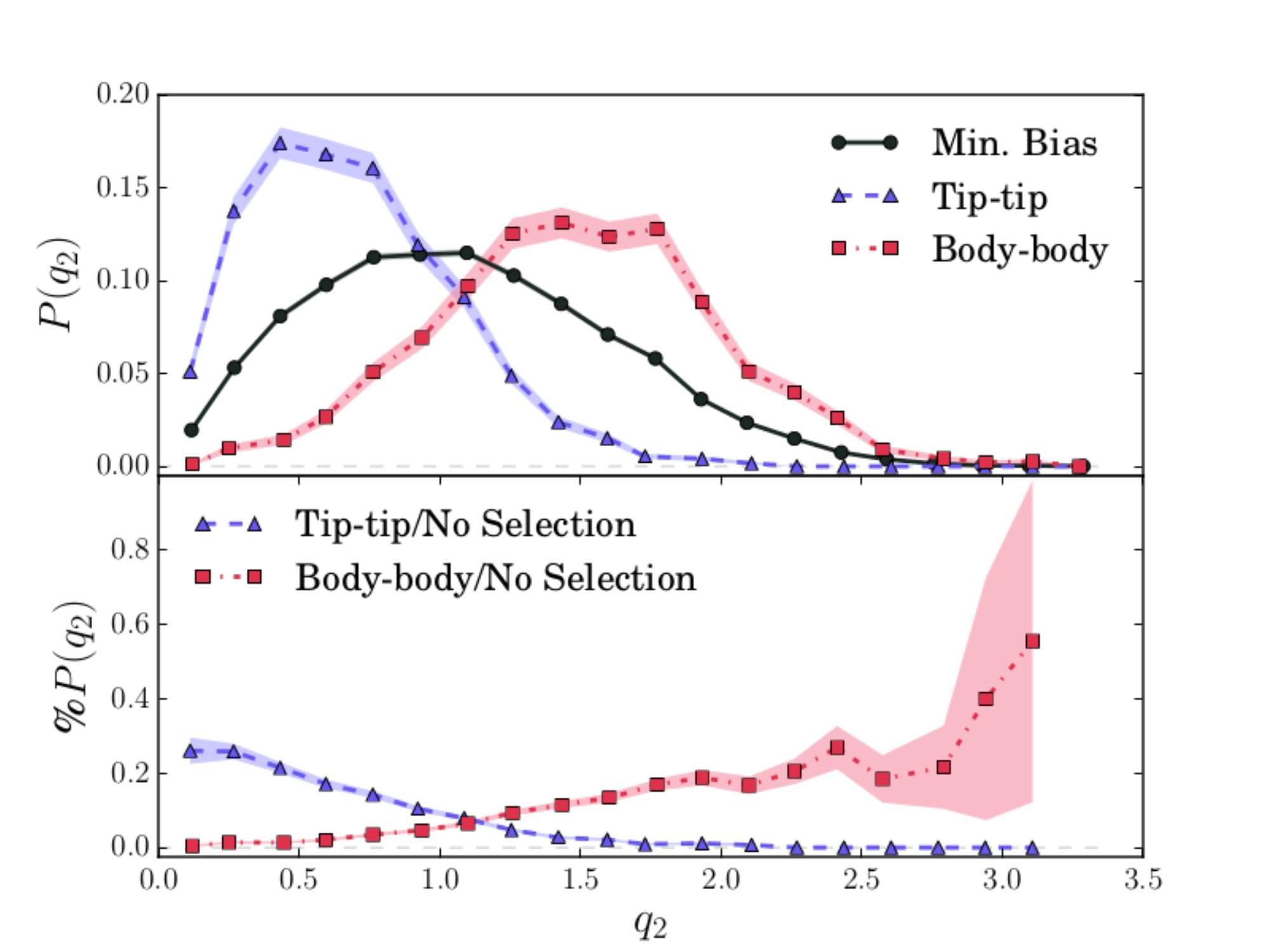}\\[-4ex]
  \subfigimg[width = \linewidth]{c)}{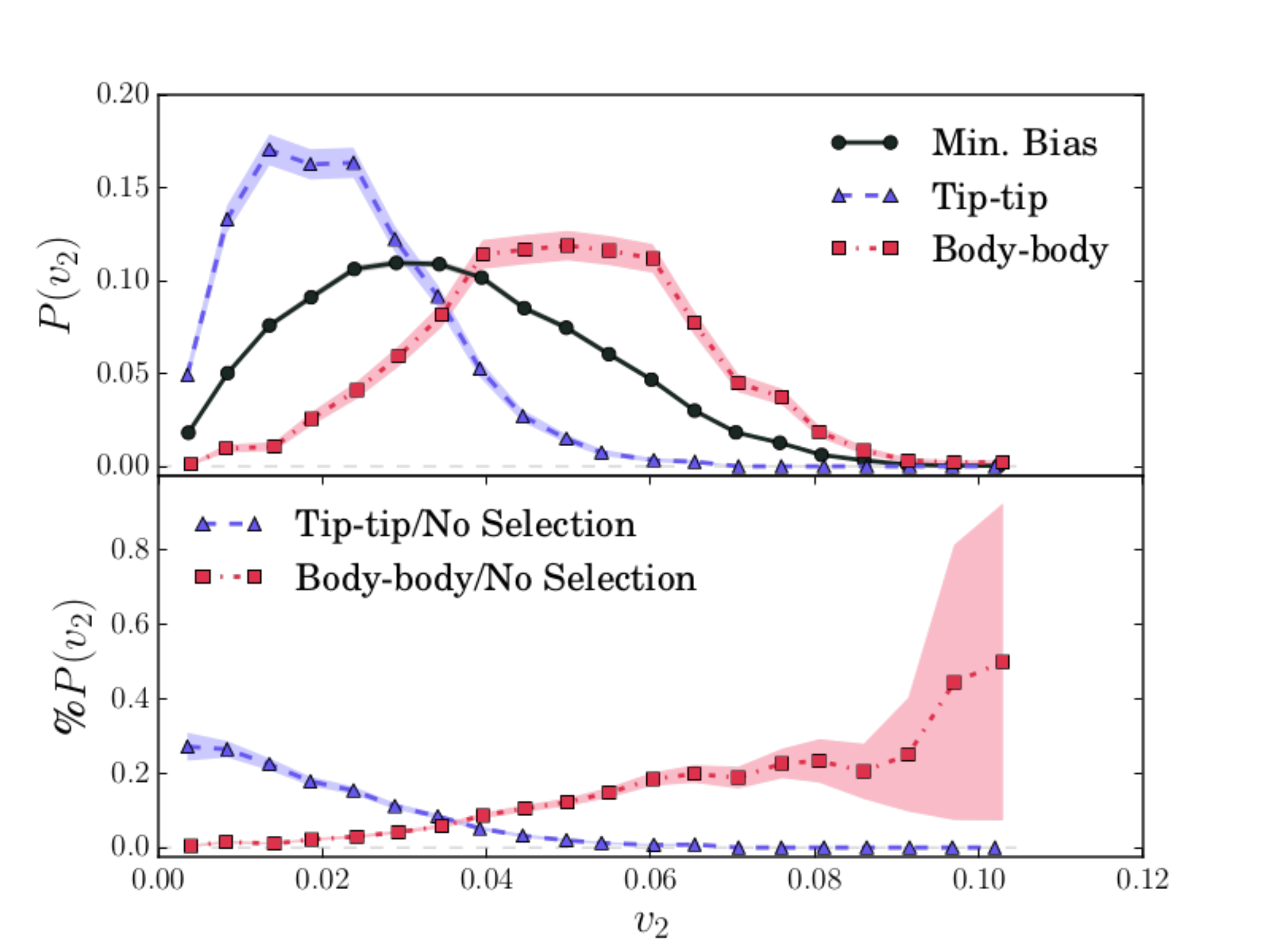}\\[-2ex]
  \caption{Probability distributions for $\varepsilon_2$ (a), $q_2$ (b) and $v_2$ (c) for different event classes within a sample of 1\% ZDC events. The top half of each panel panels shows the distributions of tip-tip and body-body collisions, while the bottom half shows the relative contributions from each class.}
  \label{fig:5}
\end{figure}

We therefore consider ``event engineering'', an approach for selecting events by the magnitudes $q_2$ and $v_2$ of their $Q_2$ or ${\cal V}_2$ flow vectors. For high multiplcity events, $Q_2 \approx \sqrt{dN_\mathrm{ch}/dy}\ {\cal V}_2$ \cite{Schukraft:2012ah}.  In Fig.~\ref{fig:5} we plot the probability distributions for $\varepsilon_2$ (a) and compare them with those for $q_2$ (b) and $v_2$ (c). Both $q_2$ and $v_2$ are computed for directly emitted pions only, assuming that $v_2^{\pi,th}\approx v_2^\mathrm{ch}$, and similar for $q_2$. Although not quite true, this approximation should cause at most a slight rescaling of the horiziontal axes in Figs.~\ref{fig:5}b,c  \cite{Qiu:2012tm}, without changing the shape of the distributions. Calculating the full resonance decay chain for all 25,000 events would have been prohibitively expensive. Comparison of panels b and c shows very similar probability distributions for $q_2$ and $v_2$, i.e. very little influence of fluctuations of the charged particle multiplicity $dN_\mathrm{ch}/dy$ on their shape.

Since tip-tip events have on average smaller ellipticities (see upper panel in Fig.~\ref{fig:5}a), selecting events with small ellipticity (or, in experiment, small $q_2$ or $v_2$) enriches the tip-tip fraction. However, in this way we will never reach more than about 25\% purity of the tip-tip sample. On the other hand, cutting the 1\% ZDC events on large $q_2$ will enrich the sample in body-body events, with a purity that can reach about 40\% for the largest $q_2$ values.

\begin{figure}
  \centering
  \includegraphics[width = 0.5\textwidth]{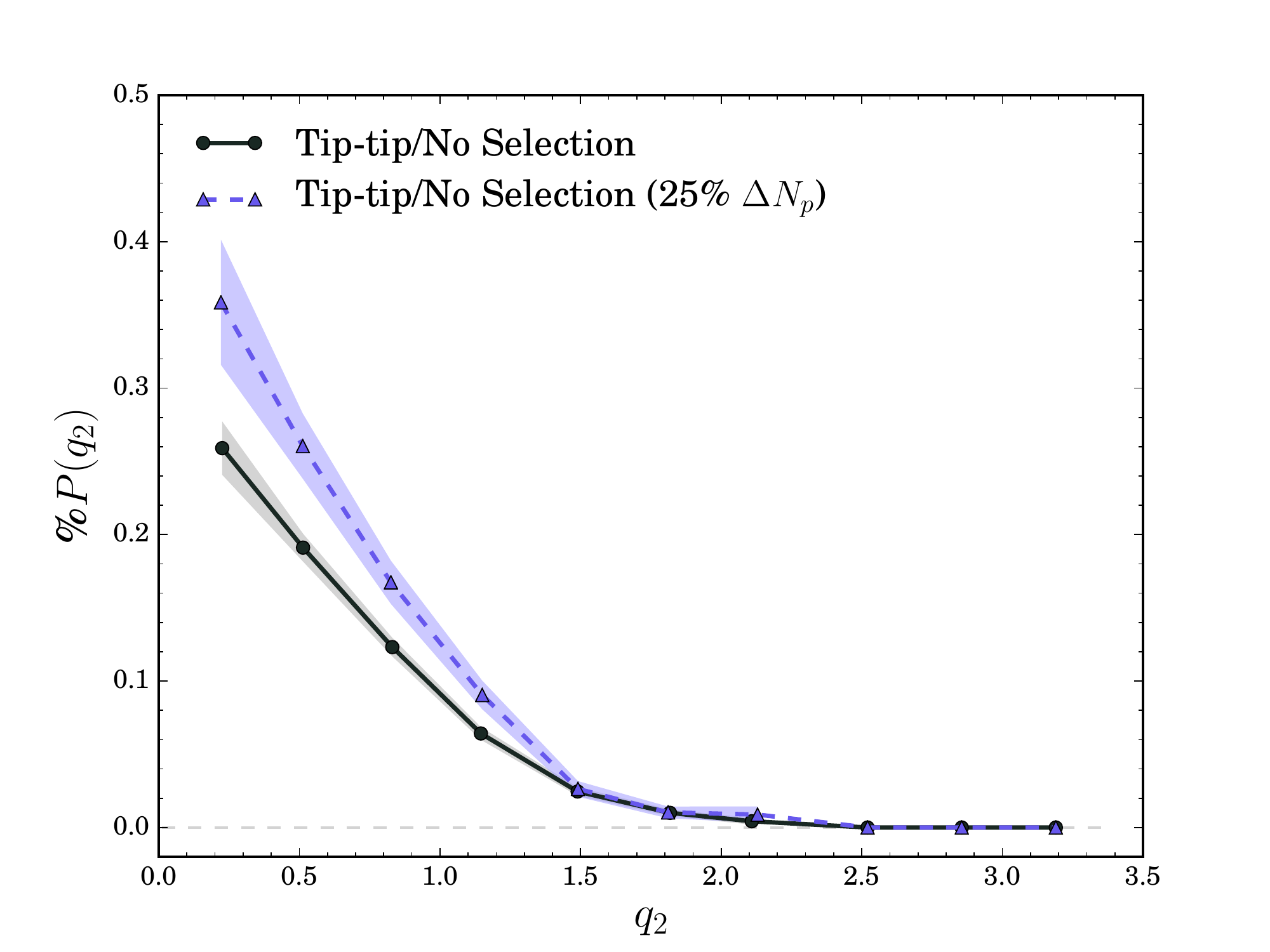}
  \caption{The solid curve shows the distribution of events scaled according to their contribution to the total probability distribution for $q_2$ as seen in the bottom right panel in Fig.~\ref{fig:5}. The dashed curve shows the increased contribution of tip-tip collosions within the 25\% of events having the smallest difference in participants $\Delta N_\mathrm{part}$ (a proxy for ZDC correlation).} 
\label{fig:6}
\end{figure}

The current ZDC cut strategy can be further refined to increase the probability of selecting tip-tip events.  Rather than looking at the ZDC signal in one of the two ZDC detectors or the sum of the ZDC signals in both detectors, we can look at the correlation of these two signals. Events with roughly equal forward and backward ZDC signals (i.e. approximately equal numbers of spectators from both nuclei) provide a better definition of the categories full overlap, tip-tip, and body-body than events with asymmetric ZDC signals where all spectators come from only one of the colliding nuclei. The difference in participants $\Delta N_\mathrm{part} = \left| N_\mathrm{part,1}-N_\mathrm{part,2} \right|$ quantifies the ZDC correlation in our model. Low values of $\Delta N_p$ correspond to the most correlated forward and backward ZDC signals. To demonstrate one application, we reconsider the purity fractions shown in the bottom part of Fig.~\ref{fig:5}b and now select from the sample only events in the lowest 25\% of $\Delta N_\mathrm{part}$. The selection on small values of $\Delta N_\mathrm{part}$ eliminates from the sample asymmetric configurations that we loosely describe as ``tip-body''.  Collisions of this type produce low values of $\varepsilon_2$ (hence $q_2$) without the angular criteria necessary to be considered tip-tip and therefore dilute the contribution of the true tip-tip configurations in the lower range of $\varepsilon_2$ ($q_2$). We show in Fig.~\ref{fig:6} that selecting the lowest 25\% of $\Delta N_\mathrm{part}$ increases the selection efficiency of $q_2$ for tip-tip configurations by a factor of about 1.4.  As a final comment, we point out that it might also be interesting to use ZDC correlations in the opposite way so as to select and study events with asymmetric tip-body configurations.

\section{Conclusions}
\label{sec:4}
Within the two-component MC-Glauber model for initial energy production, the prolate deformation of the uranium nucleus was shown to generate a knee in the centrality dependence of the ellipticity of the initial temperature distribution.  The knee was seen to be preserved by hydrodynamic evolution, after which it manifests itself in the centrality dependence of $v_2$.  Such a knee structure is not seen in the STAR data. This rules out the two-component MC-Glauber model for initial energy and entropy production. An enrichment of tip-tip configurations by triggering only on high-multiplicity in the U+U collisions thus does not work.  

To increase the selection capability between different collision geometries, we impose combined cuts on initial conditions using the spectators (ZDC), $dN/dy$, and $v_2$ (or $q_2$). For 1\% ZDC events, we found that we could enrich tip-tip collision geometries to about 50\% by cutting on high multiplicity within that sample, and body-body configurations to about 20\% purity by selecting low-multiplicity events. These numbers include effects from multiplicity fluctuations, but they rely on the binary collision admixture in the two-component MC-Glauber model and are thus model-dependent. Since the two-component Glauber model is experimentally disfavored by the failure to observe the ``knee'' structure predicted by that model in a plot of $v_2$ vs. $dN_\mathrm{ch}/dy$, these purity factors may not be reliable.

To eliminate the model-dependence just mentioned we also studied the efficiency of selecting different collision geometries by ``event engineering'', i.e. by cutting on $q_2$ or $v_2$. In this case events selected for high $q_2$ can enrich body-body collisions to about 40\% purity while cutting on low $q_2$ gives a tip-tip sample with about 25\% purity. The latter can be boosted to about 35\% purity by eliminating events with asymmetric ZDC signals. These results should not be sensitive to the binary collision admixture in the two-component MC-Glauber model and thus should be less model dependent.

\acknowledgments{This work was supported in part by the U.S. Department of Energy, Office of Science, Office of Nuclear Physics under Award No. DE-SC0004286.}

\end{document}